\documentclass[5p,twocolumn]{elsarticle}

\usepackage{hyperref}
\journal{Journal of \LaTeX\ Templates}
\newcommand{\sipm}{SiPM}
\newcommand{\ucell}{micro-cell}
\newcommand{\XT}{cross-talk}

\graphicspath{{Figures/}}

\begin{document}

\begin{frontmatter}

\title{Properties of large SiPM at room temperature}

\author{T. Montaruli~\footnote{Speaker at the conference} and A. Nagai}
\address{D\'epartement de Physique Nucl\'eaire et Corpusculaire, Facult\'e de Sciences, Universit\'e de Gen\`eve, 24 Quai Ernest-Ansermet, Geneva}

\begin{abstract}
We present in this paper the comparison of methods to measure parameters to characterize a large SiPM of area of about 1 cm$^2$. We also explain the challenges of operating it with continuous light and at room temperature. We present a method to compensate for the voltage drop induced by emission to continuous illumination in order to control its  response stability. This is important for applications where illumination changes with time, such as when they are employed in cameras of gamma-ray telescopes.
\end{abstract}

\begin{keyword}
\texttt{elsarticle.cls}\sep \LaTeX\sep Elsevier \sep template
\MSC[2010] 00-01\sep  99-00
\end{keyword}

\end{frontmatter}

\section{The Hamamatsu S10943-2832(X) large area \sipm{}}

The sensor used for the studies illustrated in this document ranks among the world's largest monolithic sensors.
It has been developed by the University of Geneva group in cooperation with the Hamamatsu company and is named S10943-2832(X). The work presented here is also the object of detailed papers \cite{Nagai:2018ovm,Nagai:2019yzb}.
The main characteristics of the sensor are detailed in Tab.~\ref{tab:HexSiPM}. The sensor is hexagonal with linear dimension of 10.4~mm flat-to-flat and area of around 93.6 mm$^2$. 

\begin{table}[hbt]
  \centering
  \renewcommand{\arraystretch}{1.4}
\begin{tabular}{||r|c||}
\hline
Nr. of channels & 4 \\
Cell size & 50 $\times$ 50 $\mu$m $^{2}$ \\
Nr of cells (per channel)         & 9210    \\
Fill Factor	       & 61.5\%  \\
$DCR$ (@$V_{op}$ per channel) 	& 2.8-5.6 MHz\\
$C_{\mu cell}$ (@ $V_{op}$ per channel) & 	85  fF\\
Cross-talk (@$V_{op}$  per channel)	&  10\%\\
$V_{BD}$ Temp. Coeff.  & 54 mV/C$^\circ$\\
Gain (@$V_{op}$  per channel) &	$1.49 \times10^6$\\
\hline
  \end{tabular}
\caption{S10943-2832(X) \sipm{} main characteristics at T = 25 $^{\circ}$C an $V_{op} = V_{BD} + 2.8$ V, including the dark count rate, $DCR$, and the capacitance of the $\mu$cell, $C_{\mu cell}$.}
\label{tab:HexSiPM}
\end{table}

\begin{figure}[t]
  \begin{center}
    \includegraphics[width=0.45\textwidth]{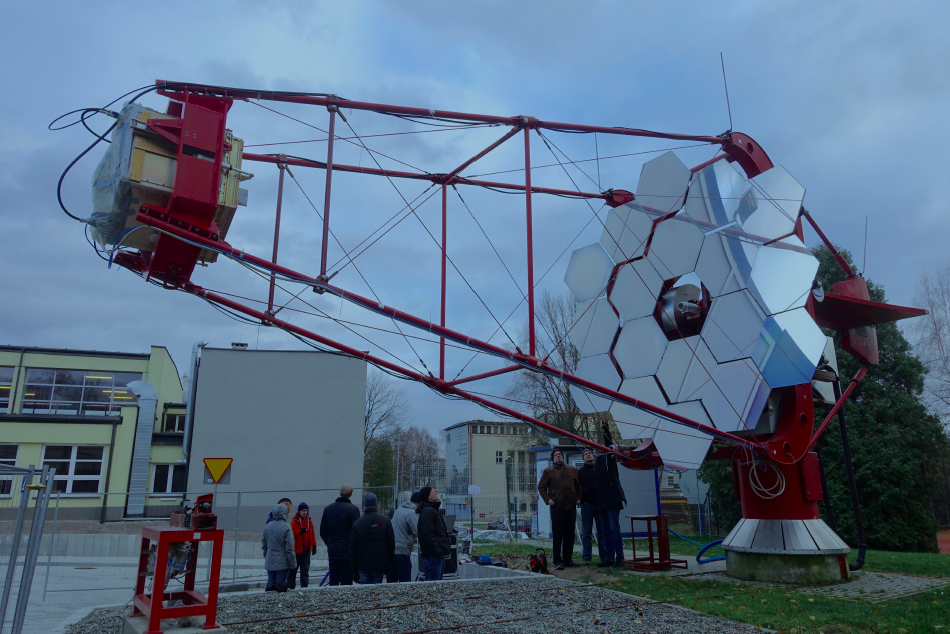}
    \includegraphics[width=0.45\textwidth]{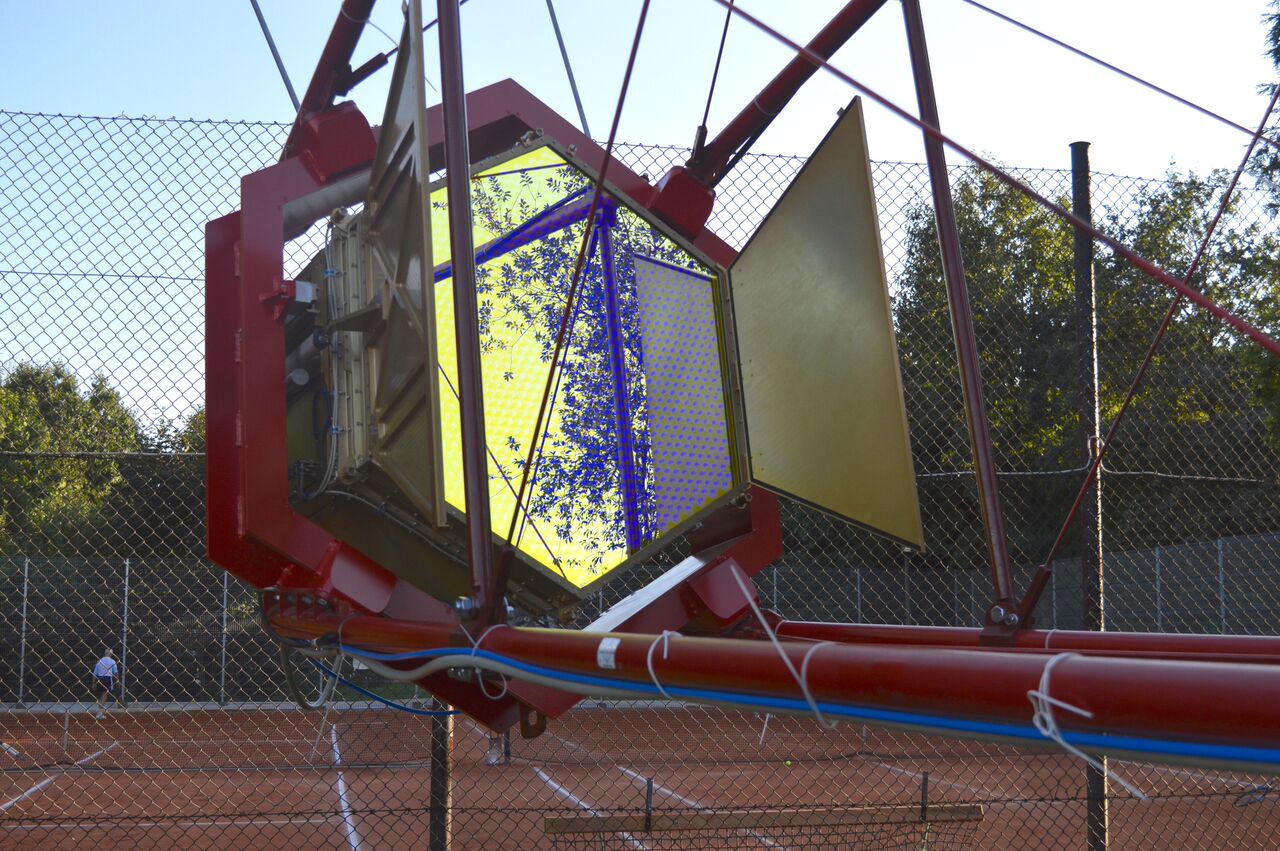}
    \caption{\label{fig:tele} 
    Top: The SST-1M telescope in Krakow since Nov. 2013.  Bottom : The camera in the telescope when it measured first light in Aug. 2017.}
  \end{center}
\end{figure}

It is currently employed in two SiPM-based cameras of the SST-1M gamma-ray telescopes (see Fig.~\ref{fig:tele} and Ref.~\cite{CameraPaperHeller2017}), that will be used in the LHAASO experiment. The telescope has been originally built for the Cherenkov Telescope Array (\href{https://www.cta-observatory.org}{CTA}) by the University of Geneva and a Consortium of Polish and Czech Institutions. 
To achieve the desired performance with the chosen optics, a mirror of 4~m diameter, the camera of the SST-1M is composed by 1296 pixels, each of an angular opening of about $0.24^\circ$. This translates into a pixel linear size of about 2.32 cm (more details on camera and its design and performances can be found in ref.~\citep{CameraPaperHeller2017}). 

\begin{figure}[t]
  \begin{center}
 \includegraphics[ width=0.7\linewidth,keepaspectratio]{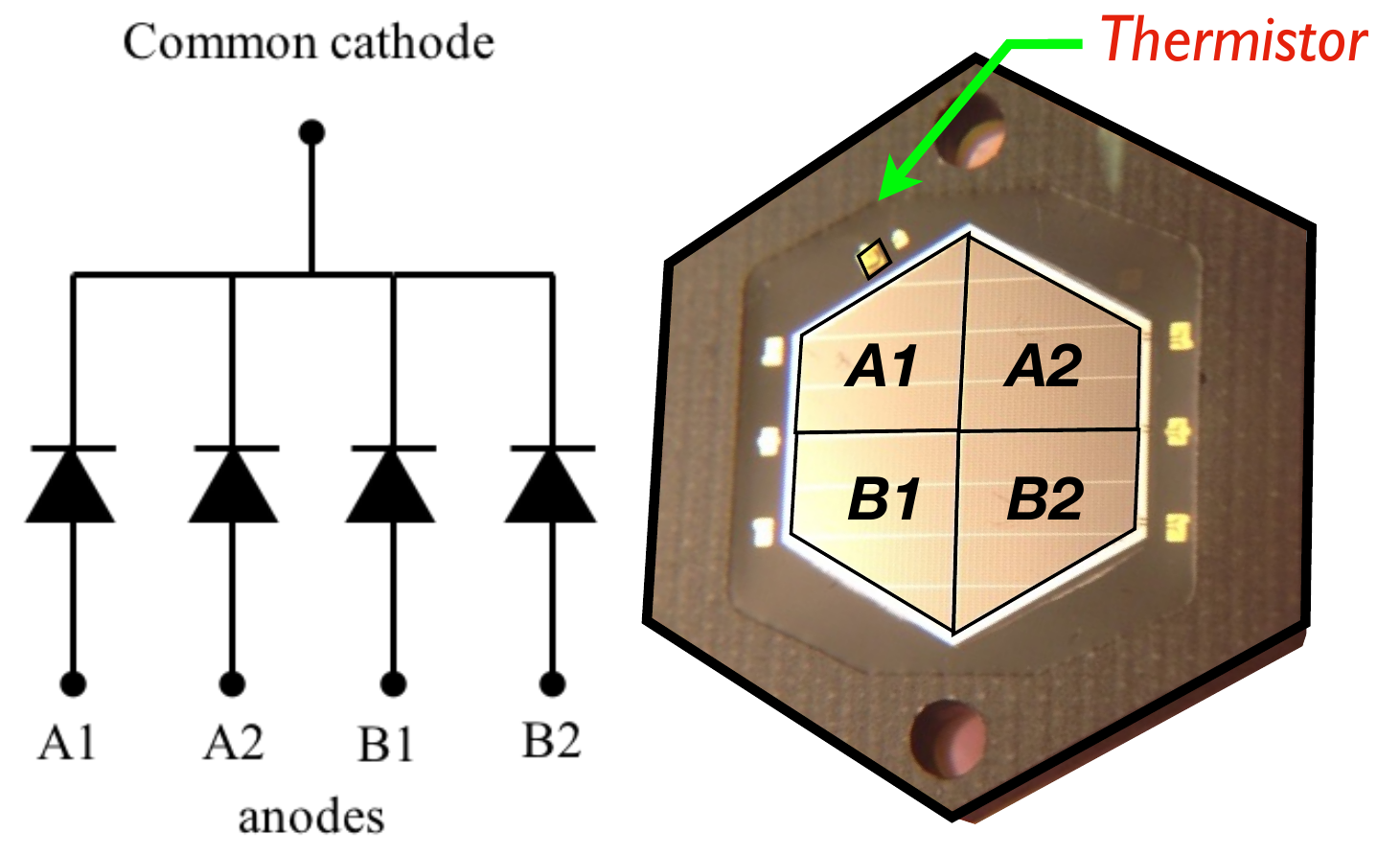}   \includegraphics[width=0.3\textwidth]{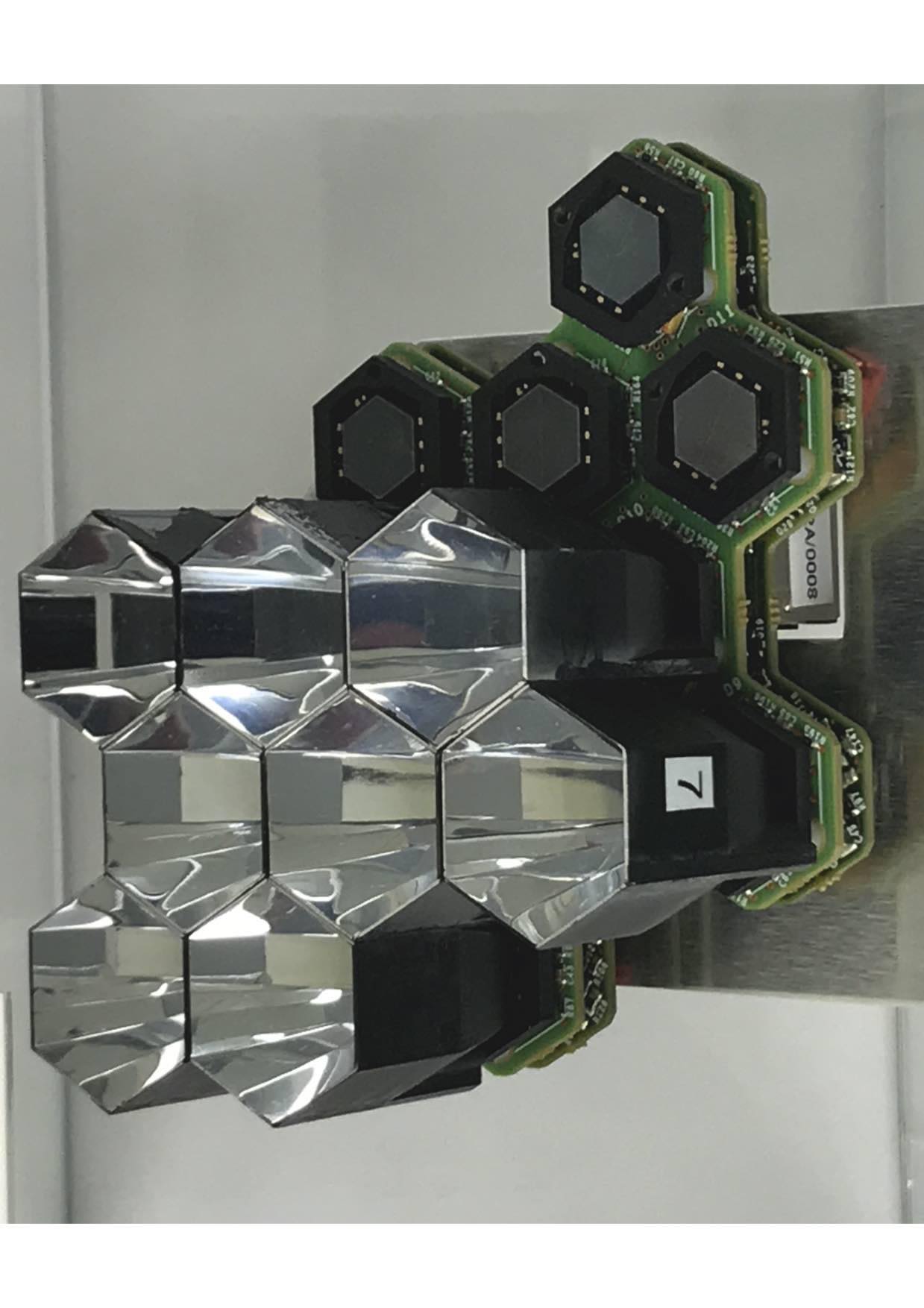}
    \caption{\label{fig:camera} 
    Top: Hamamatsu S10943-2832(X) with its electric model. The NTC probe is indicated, which monitors the temperature variations affecting SiPM working parameters~\cite{SST1Melectronics}.
    Bottom: A module of 12 pixels partially assembled.}
  \end{center}
\end{figure}

The sensor, shown in Fig.~\ref{fig:camera}-top is built with the LCT2 (Low Cross Talk) standard technology of Hamamatsu \cite{HamamatsuBook} with microcell ($\mu$cell) square size of 50 $\mu$m. Hamamatsu has further improved this technology (LCT5) and introduced the LVR (Low Reverse Voltage), which has much reduced \XT{} (XT) and increased photodetection efficiency (PDE), but  substantially longer pulse shape and higher gain. 

The hexagon is the best possible shape to achieve a uniform trigger response with minimum dead space, thanks to the fact that the pixel centres are equidistant.
The pixel size to achieve the required angular resolution is achieved through this large SiPM coupled with light funnel. 
The light funnel, approaching the ideal Winston cone geometry, has been designed by the 
University of Geneva group to be coupled to \sipm{}s and achieve the desired pixel size.
The light funnel has entrance and exit hexagonal area and has a compression factor of about six~\cite{Aguilar:WinstonCones_2014}. Its internal surface is coated in order to maximise reflection of UV Cherenkov light produced by the cosmic rays when traversing the atmosphere, and also to have a good reflectivity for light with a direction almost parallel to cone surface. A module of 12 pixels with the frontend electronics and the slow control board, replicated 108 times, makes up the full camera shown in Fig.~\ref{fig:camera}-middle. This has linear dimension of about 1~m. It is shown partially assembled in Fig.~\ref{fig:camera}-bottom.

Remarkably when using SiPM it is necessary to adopt filters reducing the contribution of the night background at waveglenths beyond $\sim 550$~nm. The SST-1M adopted a coated window of 3.3~mm thickness of borofloat with an anti-reflectant in the inner face and a filter rejecting wavelengths beyond 540~nm. 

The large area of the SiPM sensors can be a limiting factor in many applications, due to the large capacitance and dark-count rate ($DCR$). Large devices tend to have longer output signals and be more noisy.  
However, as shown by the SST-1M camera~\cite{CameraPaperHeller2017}, with the proper associated electronics, such a large device can achieve the desired performances in specific applications.
The sensor capacitance is directly related to its active area and this has an impact on the signal recharge time. 
In this case, signals would have typical duration of about hundred ns, a too  long time for the desired bandwidth of 250 MHz. 
This readout frequency of the fully digitizing electronics has been chosen taking into account the typical time duration of atmospheric showers induced by gamma-rays and cosmic rays.

To reduce the effect of the capacitance, the sensor has four independent anodes and a common cathode as shown in Fig.~\ref{fig:camera}-top.
This configuration allows to readout the 4 channels independently, but there is a single bias for the whole sensor.
Nonetheless, in order to achieve the desired bandwidth, a shaping of the signal is needed. The four channels are summed by two in order to reduce the equivalent capacitance and pulse length. The summed signals are further summed up in a differential amplifier, which feeds the output signal into the digital readout system. 
The adopted solution~\cite{SST1Melectronics} is a transimpedance amplifier topology with low noise amplifiers (\href{http://www.ti.com/lit/ds/symlink/opa846.pdf}{OPA846}) as it can achieve the required event rate with the best signal-to-noise ratio and gain/bandwidth ratio. 

\begin{figure}
  \begin{center}
  \includegraphics[width=0.45\textwidth]{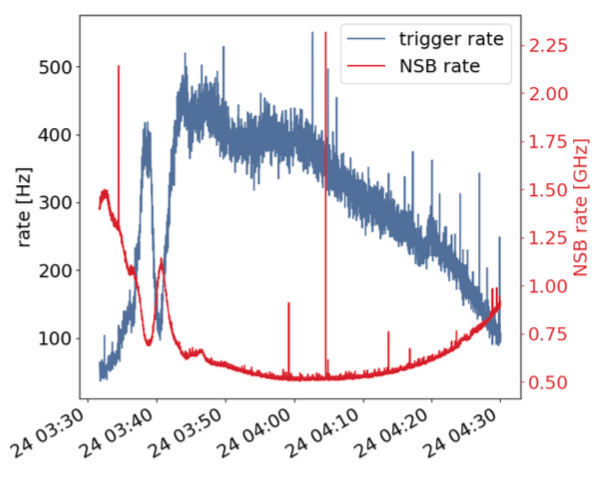}
    \caption{\label{fig:NSB} NSB and trigger rate measured by the SST-1M camera. Spikes are due to airplanes, being the telescope in Krakov close to the airport.}
  \end{center}
\end{figure}

Another important characteristic of the camera architecture is the fact that the front-end and the digital readout are DC coupled. This is important to easily measure on an event by event basis the level of Night Sky Background (NSB), meaning the moonlight and human-induced light background, which changes with time. As a matter of fact, the variation of the NSB change the real working point of the device, relevant to correctly extract the number of photons from the signal. An example of NSB variation measured by the SST-1M camera is shown in Fig.~\ref{fig:NSB}.

\section{Characterisation of the sensor}

All the laboratory measurements (i.e. static, dynamic and optical) are performed at room temperature T = 25 $^{\circ}$C at the premises of IdeaSquare\footnote{\url{http://ideasquare.web.cern.ch}} at CERN, where an experimental setup has been installed.

\subsection{Static characterisation}

The static characterization (i.e. reverse and forward current-voltage (IV) curves), is performed using a Keithley 2400~\cite{Keithley2400} pico-ammeter for bias supply and current measurements. Static means that a constant current is read. The advantage of these measurements is that they are simple and fast, while the disadvantage is that they provide only limited information, namely the breakdown voltage $V_{BD}$, the working range and $R_q$.

\begin{figure}[ht]
\begin{center}\includegraphics[%
  width=8.2cm,
  keepaspectratio]{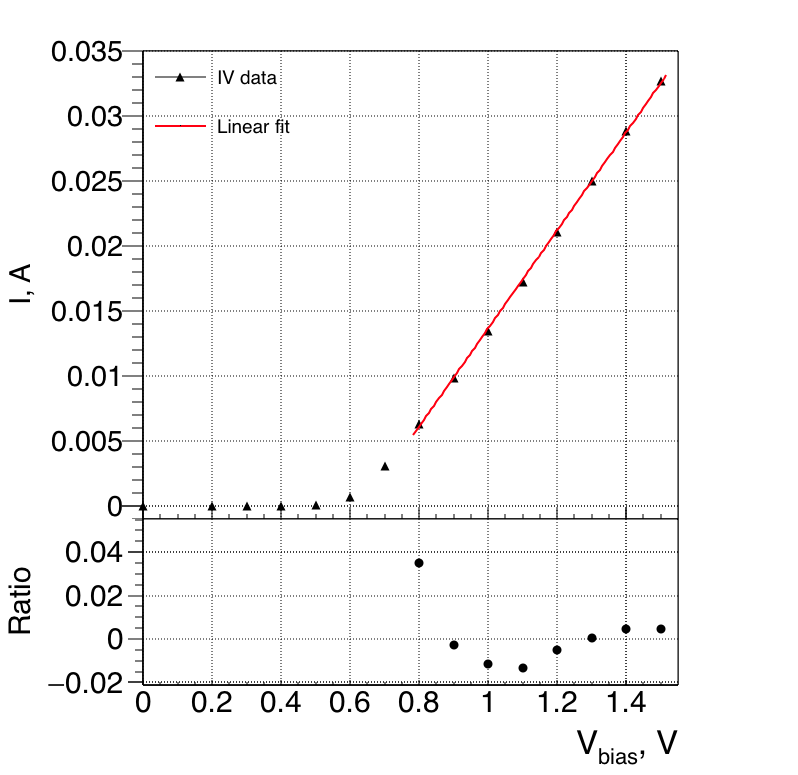}\end{center}
\caption{The forward IV characteristic and its derivative for S10943-2832(X) \sipm{}. The linear fit (red line) is superimposed to data points. In the bottom panel, the $Ratio = \left( I_{data} - I_{fit} \right)/ I_{data}$ is shown.}
\label{fig:IVForward}
\end{figure}

\paragraph{Forward IV:} As shown in Ref.~\citep{Nagai:2019yzb} the forward IV (See Fig.~\ref{fig:IVForward}) can be used to calculate the quenching resistor $R_{q}$ as:
\begin{equation}
    R_{q} = \frac{N_{\mu cell}}{b},
\end{equation}{}
where $N_{\mu cell}$ number of SiPM \ucell{}s and $b$ is the slope parameter extracted by the linear fit (red line in Fig.~\ref{fig:IVForward}). For this \sipm{}, the fit gives  
$R_{q} = 182.9 \pm 0.3$ (stat.) $\pm 31$ (sys.)  k$\Omega$. 

\begin{figure}
\begin{center}\includegraphics[ width=7.5cm, keepaspectratio]{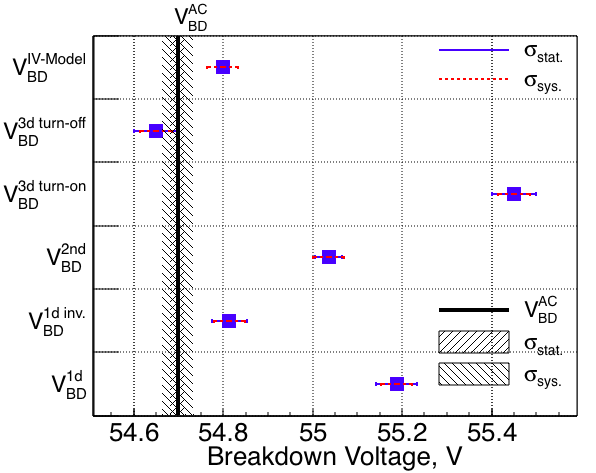}
\end{center}
\caption{$V_{BD}$ with statistics ($\sigma_{stat.}$) and systematic ($\sigma_{sys.}$) errors of the Hamamatsu S10943-2832(X) \sipm{} obtained form static and dynamic measurements.}
\label{fig:VbdComparison}
\end{figure}

\paragraph{Reverse IV:} The reverse IV measurements is commonly used for fast calculation of breakdown voltage $V_{BD}$ and \sipm{} working range. Different methods might be used to calculate $V_{BD}$, such as the ``relative logarithmic derivative''~\cite{HamamatsuBook}, the ``inverse logarithmic derivative''~\cite{InvRelDerivativeMethod}, the ``second logarithmic derivative''~\cite{2ndDerivativeMethod}, the ``third derivative''~\cite{3rdDerivativeMethod} and ``IV Model'' methods~\cite{IVModeleMethod1,IVModeleMethod2}. The $V_{BD}$ calculated from various methods are presented in Fig.~\ref{fig:VbdComparison}. They are spread over a range of less than 1~V due to the model assumptions made to approximate the IV curve with simple equations.

\subsection{Dynamic characterisation}

For the dynamic measurements presented here (also called AC measurements), instead of the standard pre-amplification topology used in the real camera~\cite{SST1Melectronics}, each \sipm{} channel is connected to an operational amplifier OPA846 and readout independently. This is done to improve precision in the measurement reducing the pile up probability.
The \sipm{} device is illuminated with low intensity light of different wavelengths (e.g. 405~nm, 420~nm, 470~nm, 505~nm, 530~nm and 572~nm) produced by pulsed LEDs.
For each operating voltage of the LED providing a certain light level, 10'000 waveforms are acquired on an oscilloscope and sampled at 500 MHz. Each one is 10~$\rm\mu$s long.  
The signal used to pulse LEDs is produced by a pulse generator and it is also used to trigger waveform acquisition. 

The readout window is adjusted in such a way to have the trigger signal in the middle of the waveform. i.e. at 5~$\rm\mu$s from the window start, in order to have: 
\begin{itemize}
\item{} a ``\textit{Dark}'' interval from 0 to 5~$\rm\mu$s, when the device is operated in dark conditions. Only uncorrelated $DCR$ enhanced by correlated noise, i.e. \XT{} (prompt and delayed) and afterpulses, are present;
\item{} ``\textit{LED}'' interval, from 5 to 10~$\rm\mu$s, when the device is illuminated by LED light pulses. In this case, both signals pulses due to the light and uncorrelated \sipm{} noise pulses are present. 
Both types of pulses are further affected by \sipm{} correlated noise (i.e. prompt and delayed \XT{} and afterpulses).
\end{itemize}
\textit{Dark} intervals  are used to calculate the \sipm{} $Gain$, the breakdown voltage $V_{BD}^{AC}$, the dark count rate ($DCR$) and the optical \XT{} probability $P_{XT}$, while \textit{LED} intervals are used to calculate the \sipm{} photon detection efficiency $PDE$. To measure the afterpulse probability, $P_{AP}$, an additional data run is performed.

The data acquisition system used for these measurements, consists of a transimpedance amplifier based on OPA846, an oscilloscope Lecroy 620Zi for the waveform acquisition (a bandwidth of 20~MHz is used to reduce the influence of the electronic noise) and a  Keithley 6487 to provide bias voltage to the \sipm{}. For each LED of different wavelengths, the over-voltage $\Delta V = V_{bias} \ - V_{BD}^{AC}$ is varied in the range \mbox{1 V~$< \Delta V < 8$~V}, to cover the full working range of the device.

\begin{figure}[ht]
\begin{center}\includegraphics[%
  width=8.2cm,keepaspectratio]{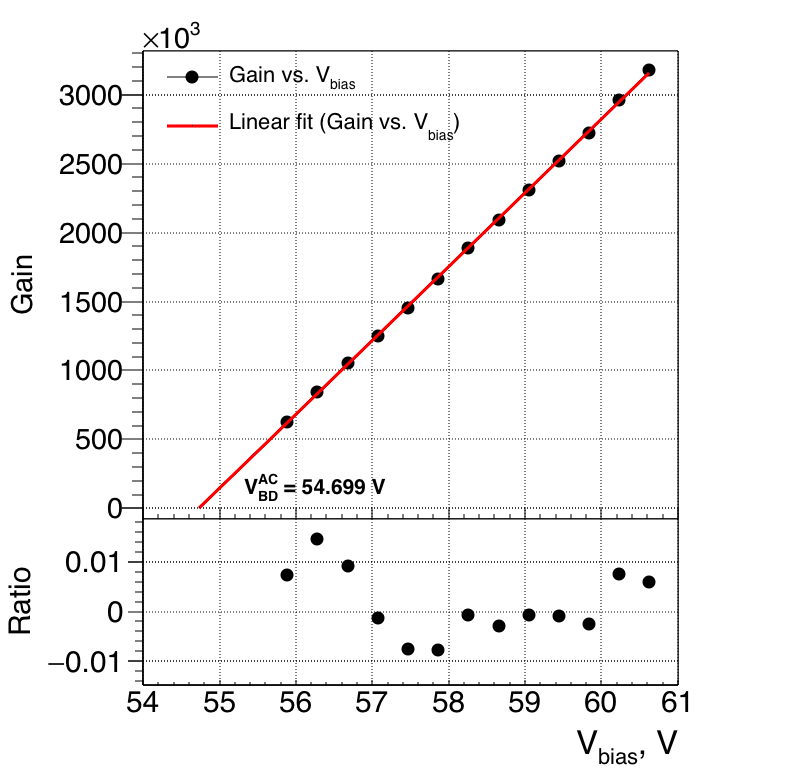}\end{center}
\caption{SiPM gain vs. $V_{bias}$ for the Hamamatsu S10943-2832(X) and ratio, defined as the difference between the experimental data and the fit function values divided by the experimental data.}
\label{fig::GanVsVbias}
\end{figure}
\paragraph{Gain and breakdown voltage $V_{BD}$:} The \sipm{} gain is calculated from the time integration of the signals of a device:
\begin{eqnarray}
 \label{Eq:GainCalc}
 G = \frac{Q}{e} = \frac{1}{G_{Amp} \cdot e}\cdot &\displaystyle \int I(t)dt
\end{eqnarray}

where $G_{Amp}$ is the amplifier gain, $I(t)$ is the current generated by \sipm{} ($I(t) = V(t)/R$). As expected, $G$ increases linearly with increasing $V_{bias}$ (see Fig.~\ref{fig::GanVsVbias}). Since no avalanches take place before breakdown, the $V_{BD}$ is calculated as the intersection of linear fits with the x-axis and  $V_{BD} = 54.699 \pm 0.017$ (stat.) $\pm$ \mbox{0.035 (sys.)}~V is found. Knowing $V_{BD}$, the overvoltage parameter, defined as $\Delta V = V_{bias} - V_{BD}$, can be determined and will be used further instead of $V_{bias}$.

\paragraph{Uncorrelated and correlated noise:}
\label{sec:Noise}

Two main categories of the \sipm{} noise can be identified: the $DCR$ or primary uncorrelated noise, which is independent from light conditions, and the secondary or correlated noise of optical \XT{} and afterpulses.
From the experimental point of view, the $DCR$ can be determined by counting all \sipm{} pulses with amplitude exceeding a given threshold (see Fig.~\ref{fig:DCRvsThreshold.png}). This \textit{counting} method is affected by correlated noise (i.e. \XT{} and afterpulses). Due to this, the \XT{} can be also calculated as:

\begin{figure}[ht]
\begin{center}\includegraphics[%
  width=0.8\linewidth,keepaspectratio]{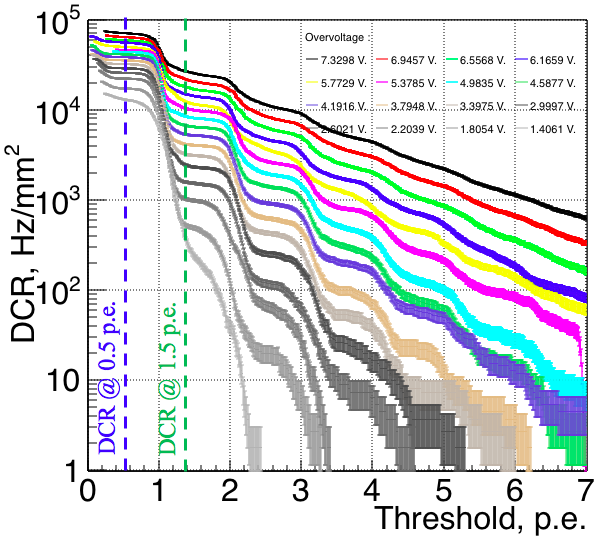}\end{center}
\caption{$DCR$ vs. threshold for different values of the overvoltage $\Delta V$ for S10943-2832(X). The blue and green vertical lines represent the $DCR$ at 0.5 p.e. and 1.5 p.e. thresholds, respectively.}
\label{fig:DCRvsThreshold.png}
\end{figure}

\begin{equation}
 \label{Eq:PXT}
 P_{XT} =  \frac{ DCR_{1.5 p.e.} } { DCR_{0.5 p.e.} } .
\end{equation}
where $DCR_{1.5 p.e.}$ and $DCR_{0.5 p.e.}$ are the $DCR$ at a threshold of 1.5 p.e. and 0.5 p.e amplitude, respectively.
To overcome the effect of correlated noise, the \textit{Poisson statistic} method can be used to calculate pure uncorrelated SiPM noise at 0.5 p.e. threshold as:
\begin{equation}
    DCR_{Poisson} = -\frac{ ln \left( P_{dark}(0) \right) }{L} =
    - \frac{1}{L}ln \left( \frac{N_{dark}(0)}{N_{dark}(total)} \right)
\label{Eq:DCRPoisson}
\end{equation}
where $P_{dark}(0)$ is the Poisson probability not to have any SiPM pulse and $-ln(P_{dark}(0))$ is the average number of detected SiPM pulses within the time interval $L$. The $P_{dark}(0)$ can be calculated as:

\begin{equation}
    P_{dark}(0) = - \frac{N_{dark}(0)}{N_{dark}(total)},
\label{Eq:DCRPoissonProb}
\end{equation}
\noindent
where $N_{dark}$(total) represents the total number of analyzed waveforms and $N_{dark}$(0) is the number of waveforms without any SiPM pulse within the given time interval $L$. As shown in Ref.~\citep{Nagai:2019yzb}, $L$ should be long enough to include all after pulses corresponding to the primary pulse, otherwise the $DCR$ would be overestimated.

\begin{figure}
\begin{center}\includegraphics[ width=8.7cm, keepaspectratio]{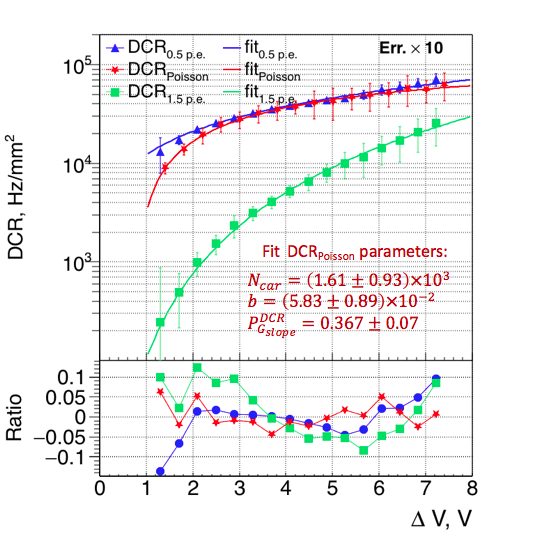}\end{center}
\caption{DCR vs. $\Delta V$ for S10943-2832(X) for the 0.5 p.e. counting method (blue), Poisson statistics  method (red) and the 1.5 p.e. threshold (green). Also shown the difference between the experimental data and the fit, normalized to experimental data, for the 0.5 (blue), 1.5 (green) p.e thresholds and Poisson statistics (red). Fit parameters for the Poisson method are given.}
\label{fig:DCRvsOvervoltage.png}
\end{figure}

The comparison of the $DCR$ calculated from the \textit{counting} and \textit{Poisson statistic} methods are presented in Fig.~\ref{fig:DCRvsOvervoltage.png}. As is expected, the \textit{counting} method shows slightly overestimated results due to afterpulses.

\begin{figure}[ht]
\begin{center}\includegraphics[%
  width=8.cm,
  keepaspectratio]{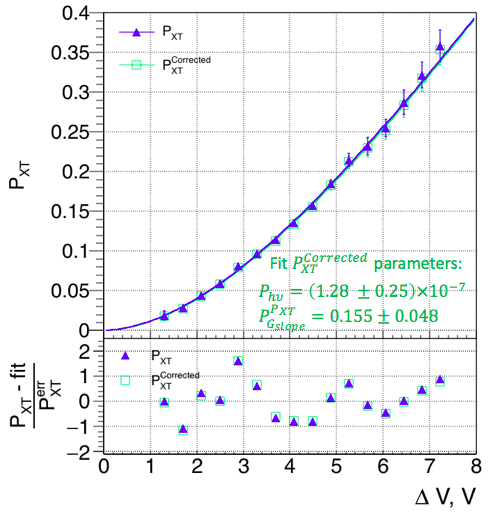}\end{center}
\caption{$P_{XT}$ vs $\Delta V$ of S10943-2832(X) with ($P^{Corrected}_{XT}$ in green) and without correction for the pile up effect ($P_{XT}$ in blue). In the bottom plot the difference between experimental data and fit normalized to data errors is shown.}
\label{fig:PXTCorr}
\end{figure}

From Eq.~\ref{Eq:PXT} the \XT{} $P_{XT}$ is calculated and presented in Fig.~\ref{fig:PXTCorr}. Using a standard approach~\citep{Nagai:2018ovm}, the $P_{XT}$ is corrected for the pile up effect and also presented in Fig.~\ref{fig:PXTCorr}.

The afterpulse probability is measured by acquiring 20~$\mu$s long waveforms, triggering their acquisition and using a pulse with an amplitude larger than 0.5~p.e.
This pulse, called in the following primary pulse, is adjusted to fall in the center of the waveform (i.e. at  10~$\mu$s). 
To ensure that pulses are either afterpulses related to the primary pulse or randomly generated dark pulses, waveforms without any signal within the 5~$\mu$s preceding the primary pulse are selected and analyzed in the following. Details of the measurement are in Ref.~\cite{Nagai:2018ovm}. 

\begin{figure}[ht]
\begin{center}\includegraphics[
  width=8.2cm,  keepaspectratio]{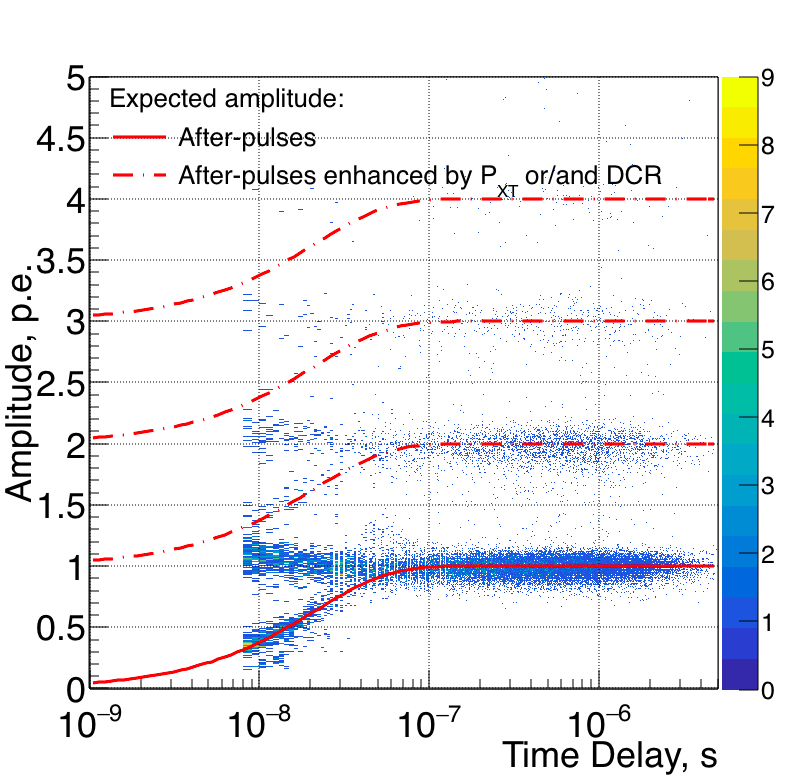}\end{center}
\caption{The 2D histogram shows the time difference between primary pulses and following ones on the x-axis and the amplitude of the second signal on the y-axis. The colors represent the number of events in each bin. The expected afterpulse amplitudes as a function of the delay time is calculated from the $\mu$cell recovery time for pure afterpulses (red solid line) and for enhanced afterpulses by optical cross-talk or dark pulses (dashed red lines).}
\label{Fig:Pap2d}
\end{figure}

Fig.~\ref{Fig:Pap2d} is a two-dimensional histogram of the amplitude in p.e. of the first pulse following a primary pulse of 1 p.e. vs the time difference between the two.
This plot shows the various \sipm{} noise components. The population of dots around amplitude of 1 p.e and time delay larger than 50 ns are typically dark pulses and afterpulses. 
The population with amplitude lower than 1 p.e. and delay smaller than 50 ns are afterpulses produced when the $\mu$cell has not yet recovered. The population at time delay less 50 ns and amplitude 1 p.e. might be mostly delayed optical cross-talk, and some dark pulses or afterpulses related to avalanches happened more than 5 $\mu$s before the primary avalanche. The other populations at larger amplitude than 1 p.e. are of similar nature than what described for 1 p.e. but further enhanced by optical cross-talk. In the plot, the red solid line is calculated from:
\begin{equation}
 \label{Eq:AfterpulseAmplitude}
 A_{AP}  =  A_{1p.e.} - A_{1p.e.} \cdot \exp \left[  - \frac{ t } { \tau_{rec.} }  \right],
\end{equation}
where $A_{1p.e.}$ is the single photoelectron (p.e.) amplitude, $\tau_{rec.} = R_{q} \cdot C_{\mu cell}$ is the recovery time constant since the afterpulsing occurs in the same $\mu$cell as primary avalanche, hence its amplitude $A_{AP}$ strongly depends on the recovery state of the $\mu$cell.

\subsection{Optical characterisation}

The photon detection efficiency $PDE$ is one of the most important parameters describing the sensitivity of a \sipm{} as a function of wavelength of the incident light  $\lambda$ and the applied overvoltage $\Delta V$:
\begin{equation}
    PDE = QE(\lambda) \times \epsilon \times P_{G}(\Delta V, \lambda)
\end{equation}
where $QE(\lambda)$ is the quantum efficiency, $P_G$ is the Geiger probability, and $\epsilon$ the \ucell{} fill factor (the percentage of it that is sensitive to light). More details about each $PDE$ component can be found in the Ref.~\cite{ACERBI201916}.
To study the $PDE$, our experimental setup at \href{https://ideasquare.web.cern.ch}{IdeaSquare} at CERN is used (more details can be found in Ref.~\citep{Nagai:2018ovm}).

\paragraph{Absolute PDE:}

\begin{figure}
\begin{center}\includegraphics[%
  width=7.5cm,
  keepaspectratio]{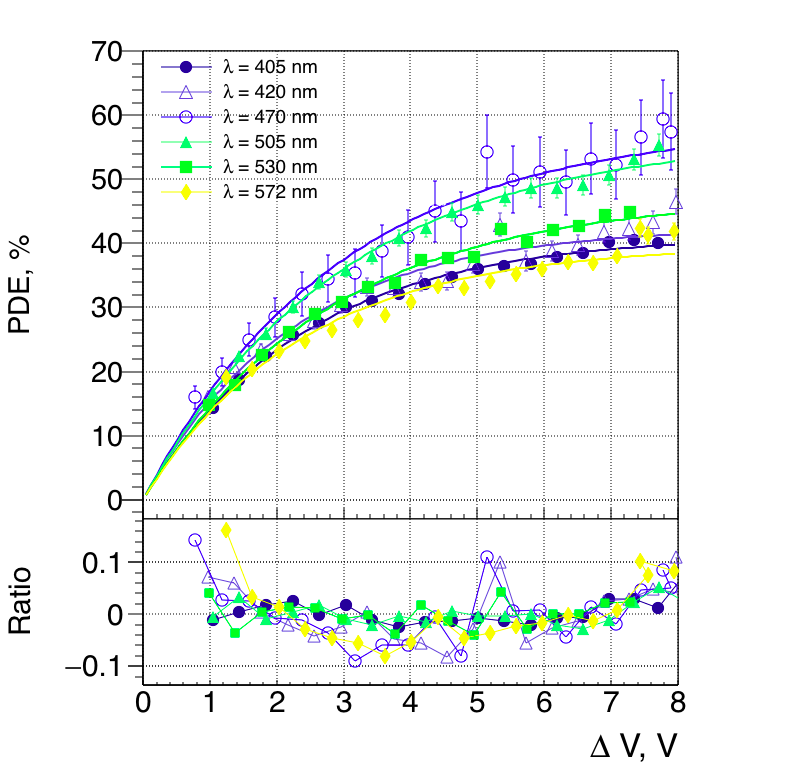}\end{center}
\caption{Absolute $PDE$ vs. $\Delta V$ of the Hamamatsu S10943-2832(X) \sipm{}. The results are presented for six different wavelengths. Also, the $Ratio = \left( PDE_{data} - PDE_{fit} \right) \div PDE_{data}$ is shown.}
\label{fig:PDEvsOvervoltage}
\end{figure}

For the absolute $PDE$ measurements, LEDs of six different wavelengths $\lambda$ are operated in a pulsed mode (pulse width from 2 to 5 ns, depending on the LEDs type). The so-called \textit{Poisson statistic} \citep{ECKERT2010217} \citep{DinuPDE} \citep{OTTE2017106} \citep{BocconeTNS} method is used for data analysis and the results are presented in Fig.~\ref{fig:PDEvsOvervoltage}. 

\paragraph{Relative PDE:}

\begin{figure}
\begin{center}\includegraphics[%
  width=7.5cm,
  keepaspectratio]{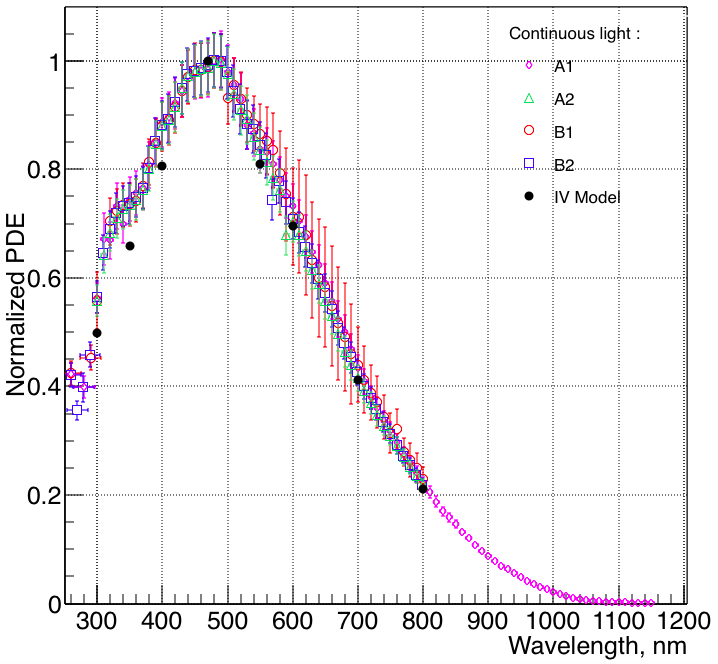}\end{center}
\caption{The relative $PDE$ vs $\lambda$ for S10943-2832(X). The results are presented for all four channels: A1, B1, A2, B2 and also for the $PDE$ calculated from the ``IV Model" for the channel A1 (black points).}
\label{fig:RelativePDEvsWavelength}
\end{figure}

The absolute $PDE$ measurements require a pulsed light source, as an LED or a laser, so it is possible only for a limited number of wavelengths. Therefore, to measure the $PDE$ in a wide wavelength range, from 260~nm up to 1150~nm, a second method, the so called ``Relative PDE'', is used. Using the method described in Ref.~\citep{Nagai:2018ovm} and \citep{DinuPDE}, the relative $PDE$ is calculated in a wide $\lambda$ range from 260~nm up to 1150~nm and presented in Fig.~\ref{fig:RelativePDEvsWavelength} for all four SiPM channels. 

\begin{figure}
\begin{center}\includegraphics[%
  width=7.5cm,
  keepaspectratio]{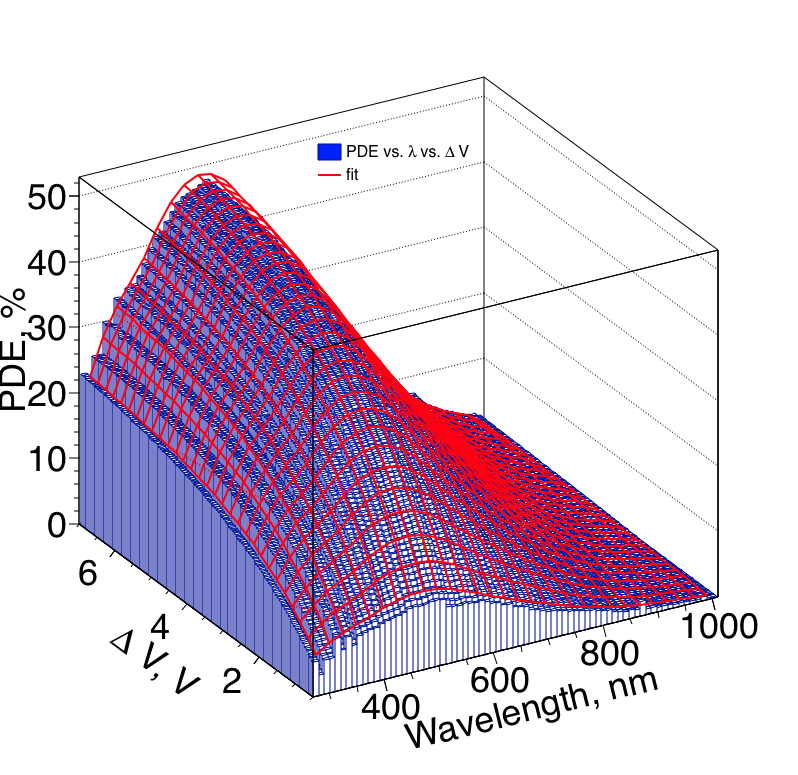}\end{center}
\caption{The PDE vs. $\lambda$ and $\Delta V$ for the SiPM S10943-2832(X).}
\label{fig:PDEvsWavelengthvsOverV}
\end{figure}

By combining the the absolute and relative $PDE$ measurements the $PDE$ as a function of $\Delta V$ and $\lambda$ can be obtained and it is presented in Fig.~\ref{fig:PDEvsWavelengthvsOverV}.

\subsection {\sipm{} \XT{} in SST-1M camera}

\begin{figure}[ht]
\begin{center}\includegraphics[%
  width=7.5cm,
  keepaspectratio]{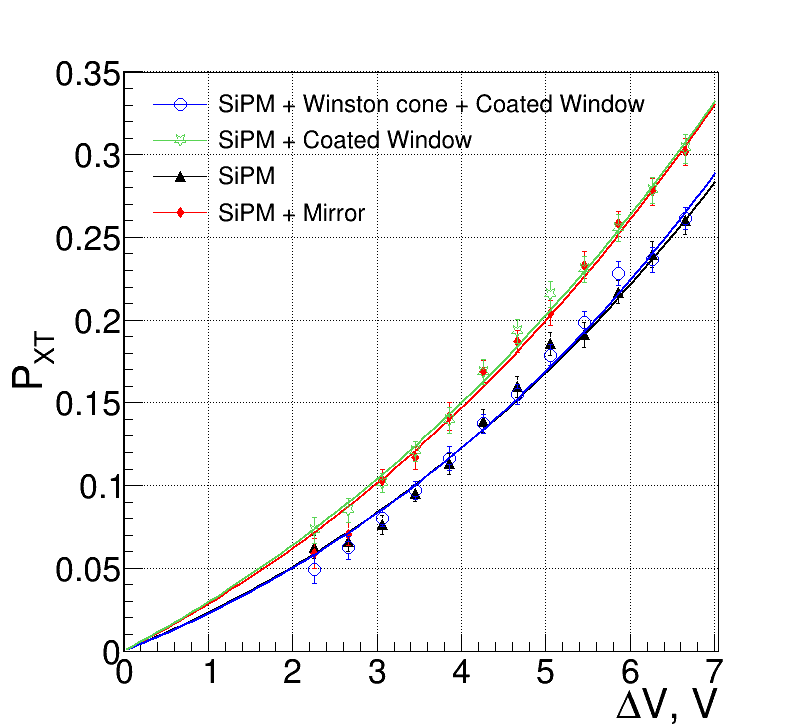}\end{center}
\caption{$P_{XT}$ vs $\Delta V$ of S10943-2832(X) measured for the \sipm{} (black), the \sipm{} covered by the coated window (green), the \sipm{} connected to the lightguide and covered by the coated window (blue) and the \sipm{} covered by the mirror (red).}
\label{fig:PXTDiffConditions}
\end{figure}

\sipm{} \XT{} is already discussed in Sec.~\ref{sec:Noise}. Different optical elements can dramatically increase the \XT{}. To study this effect, we measured the \XT{} for four different conditions: only \sipm{} device, \sipm{} covered by a reflecting mirror, \sipm{} covered by sample of a coated window used in SST-1M camera, \sipm{} connected to a coated lightguide. The results are presented in Fig.~\ref{fig:PXTDiffConditions}. As expected, the mirror and coated window (by reflecting the photons created during the avalanche multiplication process) increase the \XT{} with respect to its initial values. However, after introducing the light funnel between the \sipm{} and the coated window, the \XT{} decreases back to its initial value. This effect is obtained due to the lightguide design, which is hollow inside. Such \XT{} reduction might not be achievable with solid (i.e. quartz) light funnels, such as those adopted in the FACT camera \cite{FACT}.

\subsection{SiPM behavior under continuous light (CL)}
\begin{figure}[hbt]
    \begin{center}
        \includegraphics[width=0.52\textwidth, trim=0 2.cm 0 0, clip]{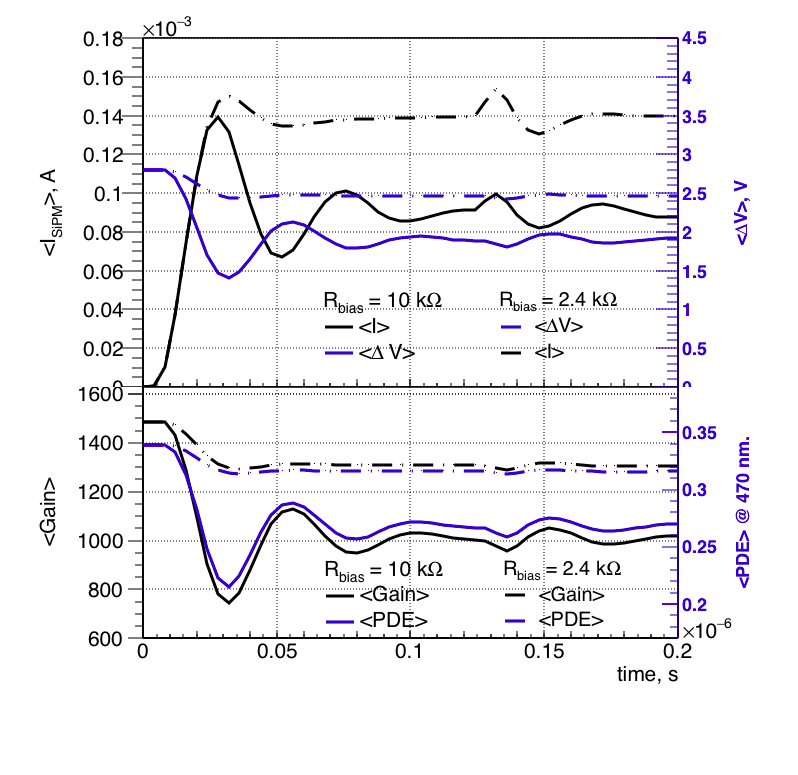}
    \end{center}
    \caption{The simulated current in a SiPM, $I_{SiPM}$, vs time under CL with photon rate of $F_{light} = 2$~GHz. $I_{SiPM}$ induces a drop of $\Delta V$, which in turn affects all main SiPM parameters (i.e. $G$, $PDE$, $P_{XT}$ , $P_{ap}$ and $DCR$). The $PDE$ and $G$ vs. time are presented in the bottom for two values of $R_{bias}$ of 10~k$\Omega$ and 2.4~k$\Omega$.}
    \label{Fig:ParamVsTime}
    \end{figure}
    
In order to protect a SiPM from drawing too high current, a resistor is connected in series to it in its bias circuit. Consequently, the resistor introduces a voltage drop when the SiPM draws a steady current. This happens when it is illuminated by constant light. This reduces the actual SiPM bias and then its sensitivity to light. As a matter of fact, this effect changes all relevant SiPM features, both electrical (i.e. breakdown voltage, gain, pulse amplitude, dark count rate and optical crosstalk) and optical (i.e. photon detection efficiency). Of course the intensity of the effect is more prominent for larger resistors and sensors with high gain (i.e. current). 

We have characterized the Hamamatsu device under light rates raging from 3~MHz up to 5~GHz of photons per sensor at room temperature (T = 25 $^{\circ}$C). 
Then, we developed a model in order to derive the parameters needed to correct for the voltage drop effect (for more details see Ref.~\citep{Nagai:2019yzb}). 
 
In Fig.~\ref{Fig:ParamVsTime} the effect on the $PDE$ and $G$ as a function of time is shown from the model for two different values of the resistor. The time interval before the steady state is achieved increases with the rate of photons $F_{light}$ on the sensors and gain $G$. For this particular example, it is reached after 100 ns and with noticeably changed parameters. 

\begin{figure}[hbt]
    \centering
    \includegraphics[width=0.5\textwidth]{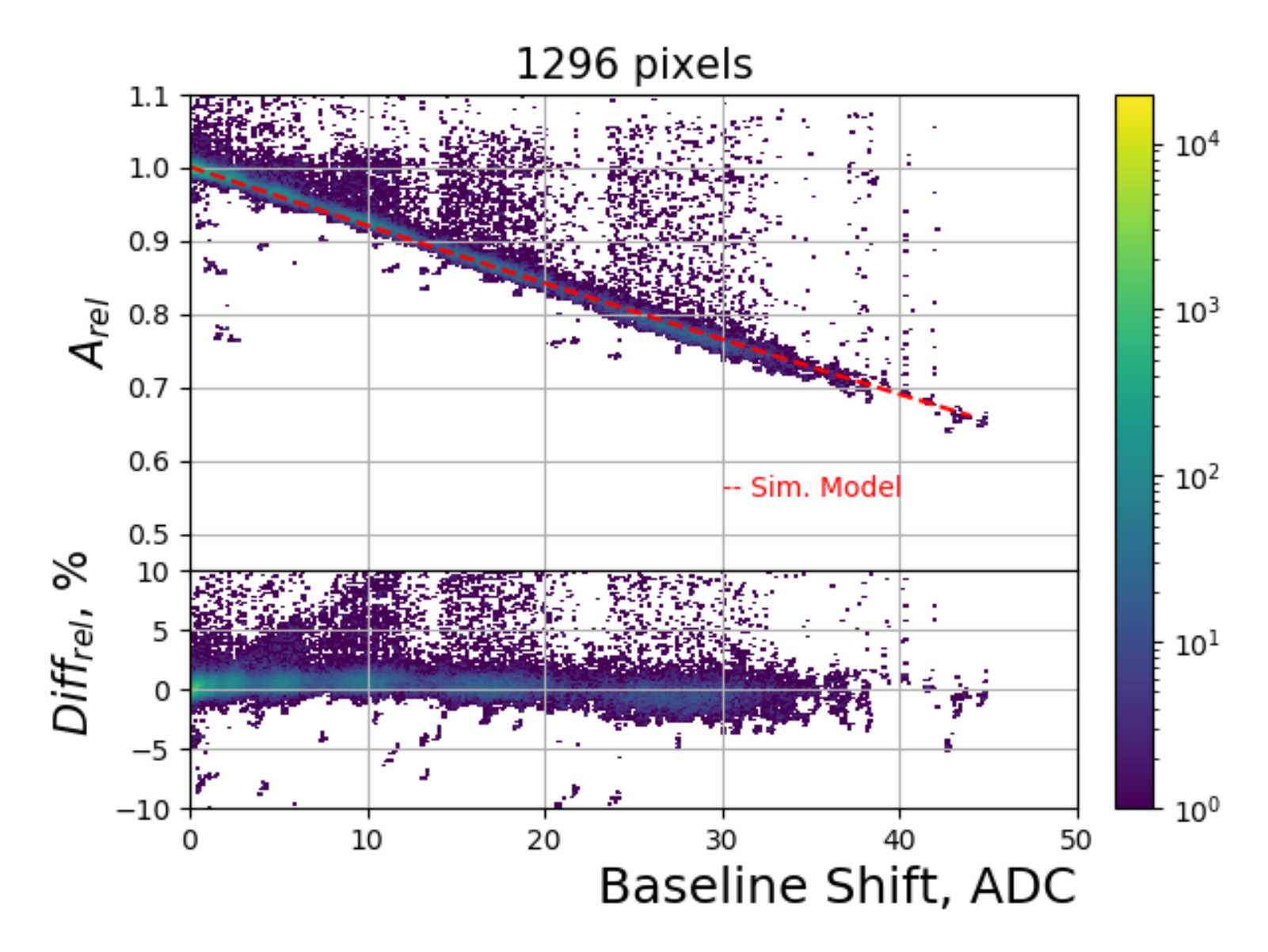}
    \caption{DC/AC scan using 12 different CL levels ($DC$ LED DAC) and injecting for each of them 19 $AC$ LED DAC level pulses. The relative amplitude $A_{rel}$ (red line from the model, points from m easurements) for all 1296 camera pixels vs. the baseline shift. The relative difference in the $A_{rel}$ for data and simulation is presented in the bottom plot. Few pixels do not follow the observed decrease mostly because of faulty LEDs or saturation of pixel readout chain~\cite{SST1Melectronics}.}
    \label{fig:AmpliDropCTS}
\end{figure}

The proposed model was implemented in a Monte Carlo and validated using the SST-1M camera and its Camera Test Setup (CTS) \cite{CameraPaperHeller2017}. The CTS is equipped with two LEDs ($\lambda$ = 468~nm) per each SST-1M camera pixel: one in pulsed mode ($AC$ LED) and the other in continuous mode ($DC$ LED). An $AC/DC$ scan, described in Ref.~\cite{Nagai:2019yzb}, has been performed for all 1296 camera pixels at $\Delta V$ = 2.8 V. For each pixel, AC and DC LED values, the baseline and signal amplitude are calculated. 
Results are presented in Fig.~\ref{fig:AmpliDropCTS}. Here the variation of the average amplitude $A_{rel.}$ of the AC LED signal on top of the baseline due to the rate of CL $F_{light}$ is shown. We can observe that almost for all pixels $A_{rel.}$ decreases with increasing baseline shift light illumination.

\begin{figure}[hbt]
    \begin{center}
        \includegraphics[width=0.45\textwidth]{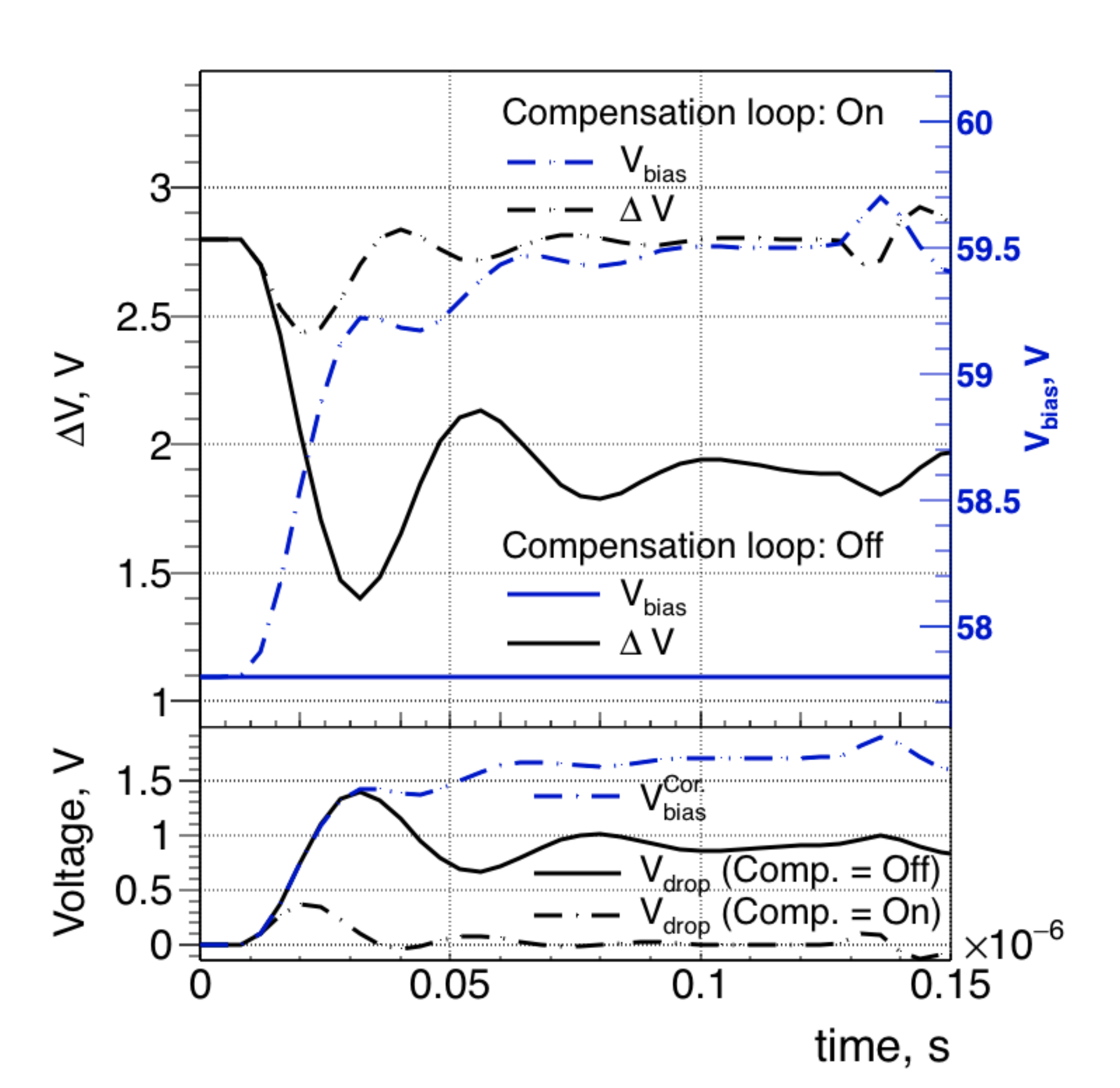}
    \end{center}
    \caption{Evolution of the over-voltage $\Delta V$ and $V_{bias}$ averaged over 10'000 samples with time under a CL of $2\times 10^{9}$ photons/s, with and without compensation loop (top). The voltage drop $V_{drop}$ and the bias voltage correction $V_{bias}^{Cor.}$ are presented at the bottom.}
\label{Fig:VoltageCorrection}
\end{figure}

The drop of the SiPM parameters under CL may be compensated by increasing the bias voltage by some correction voltage $V_{bias}^{Cor.}$ in order to keep constant the overvoltage $\Delta V$. This can be implemented in a compensation loop. The $V_{bias}^{Cor.}$ as a function of the baseline shift is presented in Fig.~\ref{Fig:VoltageCorrection} for two values of $R_{bias}$ of 10~k$\Omega$ and 2.4~k$\Omega$ with CL of $2\times 10^{9}$ photons/s and with compensation loop (dashed lines) and without (solid lines) (top). We can observe, that to compensate by $V_{drop} \sim 0.9$~V, the $V_{bias}$ should be increased by 1.7~V, as shown in Fig.~\ref{Fig:VoltageCorrection} (bottom). As a drawback, the detected NSB rate increases as well as the SiPM power consumption. 

\section{Conclusions}

We have characterized for realistic operation at an astronomical site a large area SiPM sensor. The work done can be used for any SiPM used at room temperature and in the presence of continuous light. 
\section*{Acknowledgements}
This project has received funding from the European Union's Horizon 2020 research and innovation programme under grant agreement No 713171. Also, we acknowledge the support of the SNF funding agency.

\bibliography{SST-1M}

\begin{thebibliography}{10}
\expandafter\ifx\csname url\endcsname\relax
  \def\url#1{\texttt{#1}}\fi
\expandafter\ifx\csname urlprefix\endcsname\relax\def\urlprefix{URL }\fi
\expandafter\ifx\csname href\endcsname\relax
  \def\href#1#2{#2} \def\path#1{#1}\fi

\bibitem{Nagai:2018ovm}
A.~Nagai, et~al., {\it Characterization of a large area silicon
  photomultiplier}, NIM A 948 (2019) 162796.

\bibitem{Nagai:2019yzb}
A.~Nagai, C.~Alispach, D.~Della~Volpe, M.~Heller, T.~Montaruli, S.~Njoh,
  Y.~Renier, I.~Troyano-Pujadas, {SiPM behaviour under continuous light}, JINST
  JINST\_009P\_1019 (2020) { }.
\newblock \href {http://arxiv.org/abs/1910.00348} {\path{arXiv:1910.00348}}.

\bibitem{CameraPaperHeller2017}
M.~Heller, et~al., {\it An innovative silicon photomultiplier digitizing camera
  for gamma-ray astronomy}, The European Physical Journal C 77~(1) (2017) 47.

\bibitem{HamamatsuBook}
{Hamamatsu Photonics K.K. Opto-semiconductor Handbook Editorial Committee},
  {\it Opto-semiconductor Handbook}, {Hamamatsu Photonics K.K.Solid State
  Division} (2014) 60--61.

\bibitem{Aguilar:WinstonCones_2014}
J.~Aguilar, et~al., {\it Design, optimization and characterization of the light
  concentrators of the single-mirror small size telescopes of the Cherenkov
  Telescope Array}, {Astrop. Phys.} 60 (2015) 32.

\bibitem{SST1Melectronics}
J.~Aguilar, et~al., {\it The front-end electronics and slow control of large
  area SiPM for the SST-1M camera developed for the CTA experiment}, NIM A 830
  (2016) 219 -- 232.

\bibitem{Keithley2400}
\href{http://www.testequipmenthq.com/datasheets/KEITHLEY-2400-Datasheet.pdf}{Series
  2400 sourcemeter line}.
\newline\urlprefix\url{http://www.testequipmenthq.com/datasheets/KEITHLEY-2400-Datasheet.pdf}

\bibitem{InvRelDerivativeMethod}
E.~Garutti, M.~Ramilli, C.~Xu, W.~L. Hellweg, R.~Klanner, {\it Characterization
  and X-Ray Damage of Silicon Photomultipliers}, PoS TIPP2014 (2014) 070.

\bibitem{2ndDerivativeMethod}
M.~Simonetta, et~al., {\it Test and characterisation of SiPMs for the MEGII
  high resolution Timing Counter}, {NIM A} 824 (2016) 145.

\bibitem{3rdDerivativeMethod}
{N. Ferenc, G. Hegyesi, K. Kalinka and J. Moln\'ar}, {\it A model based \{DC\}
  analysis of SiPM breakdown voltages}, {NIM A} 849 (2017) 55.

\bibitem{IVModeleMethod1}
{N. Dinu, A. Nagai and A. Para}, {\it Breakdown voltage and triggering
  probability of SiPM from IV curves at different temperatures}, NIM A 845
  (2017) 64, {Proc. of the Vienna Conf. on Instr. 2016}.

\bibitem{IVModeleMethod2}
A.~{Nagai}, N.~{Dinu}, A.~{Para}, {\it Breakdown voltage and triggering
  probability of SiPM from IV curves}, in: 2015 IEEE Nucl. Sci. Symp. and Med.
  Im. Conf., 2015, pp. 1--4.

\bibitem{ACERBI201916}
F.~Acerbi, S.~Gundacker, {\it Understanding and simulating SiPMs}, NIM A 926
  (2019) 16 -- 35, silicon Photomultipliers: Technology, Characterisation and
  Applications.

\bibitem{ECKERT2010217}
P.~Eckert, H.-C. Schultz-Coulon, W.~Shen, R.~Stamen, A.~Tadday, {\it
  Characterisation studies of silicon photomultipliers}, NIM A 620~(2) (2010)
  217 -- 226.

\bibitem{DinuPDE}
V.~Chaumat, C.~Bazin, N.~Dinu, V.~Puill, J.-F. Vagnucci, {\it Absolute photo
  detection efficiency measurement of silicon photomultipliers}, in:
  Proceedings of Science, 2012.

\bibitem{OTTE2017106}
A.~N. Otte, D.~Garcia, T.~Nguyen, D.~Purushotham, {\it Characterization of
  three high efficiency and blue sensitive silicon photomultipliers}, NIM A 846
  (2017) 106 -- 125.

\bibitem{BocconeTNS}
{A. Basili, J.A. Aguilar, A. Christov, D. della Volpe, T. Montaruli, M.
  Rameez}, {\it Characterization of New Hexagonal Large Area MPPCs}, {IEEE
  Trans. on Nucl. Sci.} 61 (2014) {1474}.

\bibitem{FACT}
H.~Anderhub, et~al., {\it Design and Operation of FACT -- The First G-APD
  Cherenkov Telescope}, JINST 8~(06) (2013) P06008--P06008.

\end{thebibliography}

\end{document}